# Charge order suppression and antiferromagnetic to ferromagnetic switch over in $Pr_{0.5}Ca_{0.5}MnO_3$ nanowires

S.S.Rao, K.N.Anuradha, S.Sarangi and S.V.Bhat[a)]

*Department of Physics, Indian Institute of Science, Bangalore, 560012, India*

We have prepared crystalline nanowires (diameter ~ 50 nm, length ~ a few microns) of the charge ordering manganite $Pr_{0.5}Ca_{0.5}MnO_3$ using a low reaction temperature hydrothermal method and characterized them using X-ray diffraction, transmission electron microscopy, SQUID magnetometry and electron magnetic resonance measurements. While the bulk sample shows a charge ordering transition at 245 K and an antiferromagnetic transition at 175 K, SQUID magnetometry and electron magnetic resonance experiments reveal that in the nanowires phase, a ferromagnetic transition occurs at ~ 105 K. Further, the antiferromagnetic transition disappears and the charge ordering transition is suppressed. This result is particularly significant since the charge order in Pr0.5Ca0.5MnO3 is known to be very robust, magnetic fields as high as 27 T being needed to melt it.

[a)]Electronic mail: svbhat@physics.iisc.ernet.in

The competition between and the coexistence of charge ordering (CO) and colossal magneto resistive (CMR) ferromagnetic (FM) ground states in the rare earth manganites $R_{1-x}$ $A_x$ $MnO_3$ (where R is a trivalent rare earth ion and A is a divalent alkaline earth ion) is currently the subject of intense theoretical[1] and experimental[2] study. It is generally believed that the two phases are mutually exclusive; e.g. in the prototype manganite $La_{1-x}Ca_xMnO_3$ for $0.2 < x < 0.5$ the material is ferromagnetic and metallic while for $0.5 < x < 0.9$ it is antiferromagnetic and charge ordered. Two essentially different mechanisms have been proposed for these two apparently disparate phenomena, namely, Zener double exchange for FM, metallic phase and polaron hopping for the CO, AFM phase. However, recently a number of experiments have provided evidence[2,3] for the coexistence of the two phases under certain circumstances, as well as for the fragility of the phase boundaries between them. It was observed that minute perturbations of the system with magnetic fields of the order of a few Tesla, electric field and radiation could cause a "melting" of the CO phase and the appearance of an FM phase[4]. However, the CO phase in $Pr_{0.5}Ca_{0.5}MnO_3$ (PCMO:0.5) has been found to be particularly robust, magnetic fields of the order of 27 T being necessary to cause a transition to FM metallic phase[5]. In this work we show that for this system the charge order can be suppressed and FM phase be observed when the material is prepared in the form of nanowires. This result has important implications for nano-device applications of manganites and to the best of our knowledge, is the first report of charge order suppression in a one dimensional manganite system.

The nanowires of PCMO (x=0.5) were prepared using the hydrothermal method[6-8]. High purity $Pr(NO_3)_3 . 6H_2O$, $Ca(NO_3)_2. 4H_2O$ , $KMnO_4$, $MnCl_2.4H_2O$ and KOH obtained from Sigma Aldrich were used for the preparation. Stoichiometric proportions of the chemicals were dissolved in deionized water. KOH was added to adjust the alkalinity. The solution was vigorously stirred and poured into a teflon vessel. The vessel was placed in a sealed stainless steel container, which in turn was heated at $270^{o}$ C for 50 hrs in a furnace. After this autoclave was cooled and depressurized, the product of the reaction was washed with deionized water and dried in ambient air at $100^{o}$ C. Energy dispersive X-ray analysis (EDAX), showed that the compound is stoichiometric to 1%. The material was further characterized by various techniques such as X-ray diffraction, transmission electron microscopy (TEM), selective area electron diffraction (SAED), high-resolution electron microscopy (HREM), SQUID magnetometry and electron magnetic resonance (EMR) spectroscopy.

Figure 1 shows the powder XRD pattern recorded with Cu K$\alpha$ radiation in the 2$\theta$ scan range from 10 to $100^{o}$ at the rate of $0.01^{o}$/10 s. The sample is found to be single phasic and free from impurities. The Reitveld analysis of the pattern shows that the unit cell is orthorhombic with the lattice parameters a = 5.4464 $A^{o}$, b = 7.6031 $A^{o}$, c = 5.3962 $A^{o}$ and the unit cell volume V = 223.0842 $A^{o3}$. The corresponding values for the bulk are a = 5.40428 $A^{o}$, b = 7.6127531 $A^{o}$, c = 5.39414 $A^{o}$ , and V = 221.92 $A^{o3}$(ref. 9). Thus there is found to be a small increase in the unit cell volume in nanowires, with an associated contraction along the b direction. Figure 2 shows the results of TEM investigations of the nanowires. It is seen that each nanowire is of uniform diameter along its entire length. The diameter of most of the wires is found to be ~ 50 nm though a few wires with larger diameter (~120 nm) are also seen. The lengths are of the order of a few micrometers. The SAED pattern shows that the nanowires are single crystalline. HREM lattice imaging confirms the single crystalline nature of the nanowires. The interplanar spacing is found to be 3.864 $A^o$ and we conclude that the growth direction is [101].

In figure 3 are presented the results of SQUID magnetometry carried out on the nanowires along with those on bulk PCMO:0.5 presented for comparison. The latter was prepared by solid state reaction of $Pr_6 O_{11}$, $CaCO_3$ and $MnO_2$. For the bulk sample, in conformity with the already published behavior[10], the charge ordering transition at 245 K is clearly seen as a peak in the magnetization and a shallow, broad peak marks the antiferromagnetic transition at ~175 K. (The small increase in the magnetization below ~ 50 K for the bulk sample could be attributed to ordering of the $Pr^{3+}$ ions.) In sharp contrast, for the nanowires, the CO peak is greatly suppressed and the AF peak is absent. Further, a rise in the magnetization is observed starting at T < 125 K, indicative of ferromagnetism. The inflection point given by the minimum in dM/dT vs T shown in the inset of fig. 3 gives the transition temperature $T_C$ to be 105 K.

In our earlier work[11] we have observed that the transition to the ferromagnetic state in manganites is accompanied by characteristic changes in the line shapes of EPR signals. This is the consequence of the appearance of exchange and anisotropy fields in the ferromagnetic state. With a view to obtaining additional confirmation of the transition to the ferromagnetic state in the PCMO:0.5 nanowires, we have carried out X-band EPR experiments on the nanowires dispersed dilutely in paraffin wax as a function of temperature in the range 10 K- 300 K. The results are presented in figure 4. It can be seen that down to about 130 K symmetric Lorentzian signals are observed which however become very broad and asymmetric below that temperature, in a way similar to that observed in ferromagnetic manganites. The detailed analysis of the EPR results in terms of the temperature dependence of the g factor, the linewidth and the intensities will be published elsewhere. Here we only wish to point out the additional corroboration obtained by the EPR measurements towards a ferromagnetic phase in the PCMO:0.5 nanowires.

Now we address the question of the likely mechanism of the suppression of the CO and the disaapearance of the AFM transitions and the appearance of the FM phase in the nanowire system. It is well known that significant changes of the magnetic properties are expected at the nano scale. For example as long ago as 1962, Neel[12] suggested that fine particles of AFM particles should exhibit weak ferromagnetism and/or superparamagnetism. He argued that this occurs as a result of uncompensated spins on the two sub lattices. Since then a number of examples have been reported where non-zero magnetic moments have been observed in nanosize particles of bulk antiferromagnetic materials[13]. A similar mechanism, in principle, could provide an explanation for our observation, in as much as along the two directions perpendicular to the growth axis of the nanowires, surface effects dominate over the bulk effects, which can result in uncompensated spins. However, there are indications that a different mechanism is operative in the nanowire system. First, we note that it is not just the AFM transition that is absent in the nanowires but the CO transition as well has been suppressed. The model of uncompensated spins, which explains the appearance of non-zero magnetization in AFM nanoparticles cannot fully explain the suppression of the CO state. Further, our recent experiments[14] on nano particles of PCMO:0.5 have shown that particles of size ~ 20 nm show charge ordering. Only when the size becomes as small as 10 nm an FM phase appears. In contrast, in nanowires of PCMO:0.5 with dia ~ 50 nm and length ~ a few microns we observe the suppression of CO. Therefore, we believe that a possible reason for the distinct behavior of nanowires could be the effect of the low dimensionality on the electron bandwidth which is known to have a deciding role in the phase diagram of manganites.

An important correlate of the FM phase in a large number of manganites is the insulator to metal transition occurring around Tc accompanied by CMR. To find out if the nanowires also show these phenomena, the resistivity of single nanowires as functions of temperature and magnetic field needs to be measured. Our preliminary electrode-less conductivity measurements on the nanowires of PCMO:0.5 indicate that the transition to the ferromagnetic state is accompanied by a large enhancement in the conductivity..

In summary, we have prepared nanowires of the charge ordering manganite $Pr_{0.5}Ca_{0.5}MnO_3$ using a hydrothermal method and characterized them by various techniques. The charge ordered phase is found to be greatly suppressed and antiferromagnetic phase is seen to have disappeared in the nanowires. Instead, a ferromagnetic phase is observed in this one-dimensional phase of $Pr_{0.5}Ca_{0.5}MnO_3$.

This work was supported by funding from the University Grants Commission of India. SSR acknowledges a research fellowship from the Council of Scientific and Industrial Research.

[1]G. C. Milward, M. J. Calderon and P.B. Littlewood, Nature **433**, 607 (2005)

[2] J. C. Louden, N. D. Mathur, and P.A Midgley, Nature **420,** 797 (2002)

[3]C.H Chen, and S. W. Cheong, Phys. Rev. Lett., **76,** 4042 (1996)

[4]For a review see Y. Tomioka and Y. Tokura in "Colossal Magnetoresistive Oxides" ed.: Y. Tokura, (Gordon and Breach, New York, 2000) p. 281-305.

[5]M. Tokunaga, N. Miura, Y. Tomioka and Y. Tokura, Phys. Rev. **B 57**, R9377, (1998).

[6]D. Zhu, H. Zhu and Y. Zhang , Applied Physics Letters, **80**,1634 (2002).

[7]D. Zhu, H Zhu and Y.H Zang, J. Phys. : Condens. Matter, **14**, L 519- L524 (2002).

[8]T. Zhang, C.G.Jin, T. Qian, X.L.Lu, J.M.Bai and X.G.Li, J. Mater.Chem., **14,** 2787, (2004)

[9]P. M. Woodward, T. Vogt, D. E. Cox, A. Arulraj, C. N. R. Rao, P. Karen, and A. K. Cheetham, Chem. Mater., **10**, 3652-3665 (1998).

[10]Y. Tomioka, A.Asamitsu, H. Kuwahara, Y. Moritomo and Y. Tokura, Phys. Rev. **B 53**, R1689, (1996).

[11]Janhavi P. Joshi, A. K. Sood, S.V. Bhat, Sachin Parashar, A.R. Raju and C.N.R. Rao, J. Mag. Mag. Mat, **279**, 91, (2004)

[12]L. Neel. In Low Temperature Physics, ed. by C. Dewitt, B.Dreyfus, P. D. de Gennes (Gordon and Beach, New York 1962), p.413.

[13]R. H. Kodama, S .A. Makhlout, A .E. Berkowitz: Phys. Rev. Lett. **79**, 1393, (1997)

[14]S. S. Rao and S. V. Bhat, to be published (2005)

Figure captions:

Figure 1. Observed (dots) and Rietveld fitted (continuous lines) XRD patterns of $Pr_{0.5}Ca_{0.5}MnO_3$ nanowires. ($R_w = 8.3$ % )

Figure 2. Typical TEM images (a,b) and HREM and SAED patterns (c,d) of $Pr_{0.5}Ca_{0.5}MnO_3$ nanowires. The scale bar in (a) and (b) is 100 nm and in (c) it is 5 nm.

Figure 3. Temperature dependence of magnetization for the $Pr_{0.5}Ca_{0.5}MnO_3$ nanowires (filled square) and the bulk polycrystalline sample (filled circle) at a constant magnetic field of 0.1 T. The first derivative of M w.r.t. T is shown in inset.

Figure 4. EMR signals of nanowires at different temperatures. The solid lines are experimental signals and the filled circles on the signals for 310 K and 130 K are fits to the field derivative of Lorentzian lineshape.

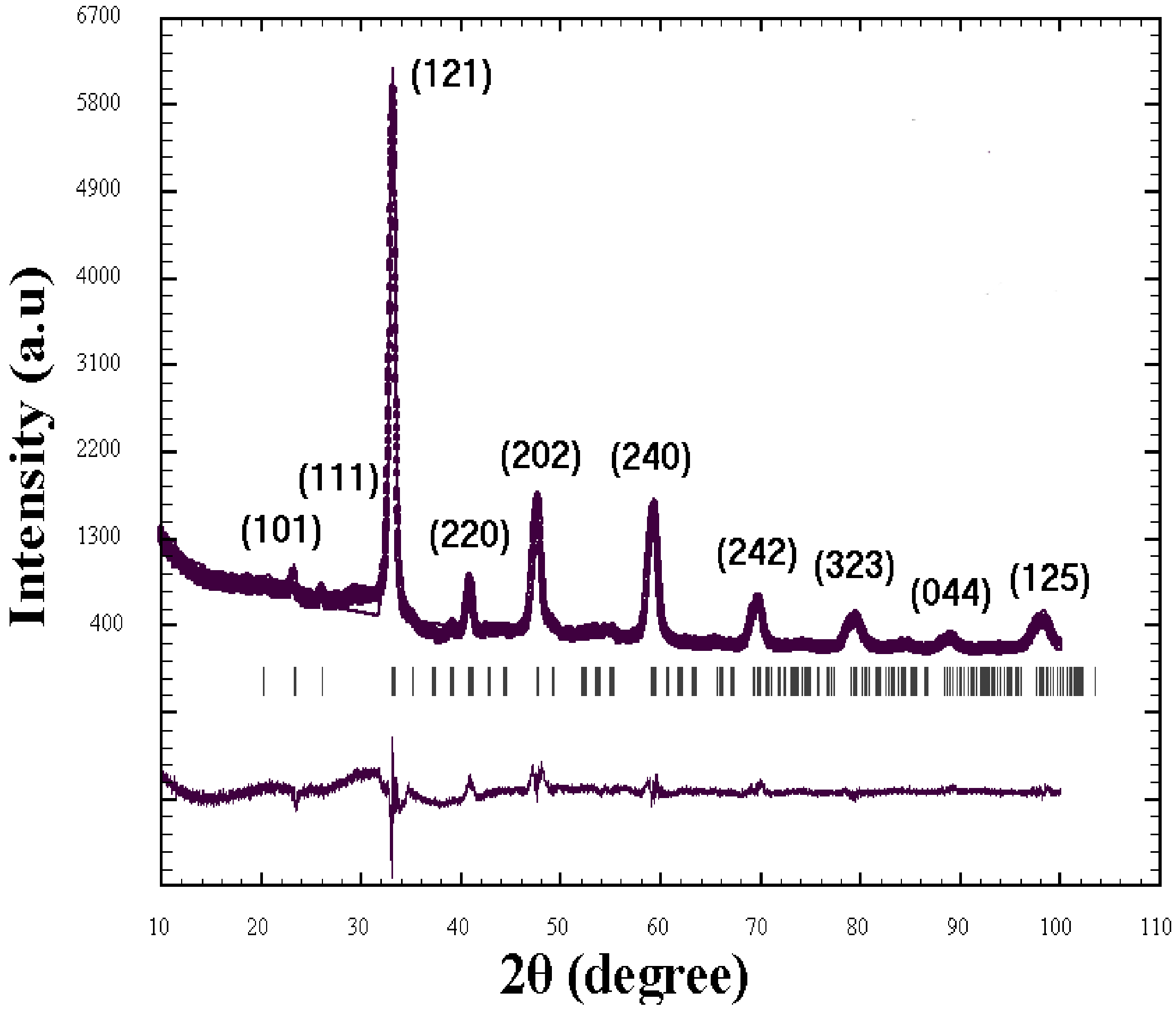

Figure 1; Rao et al

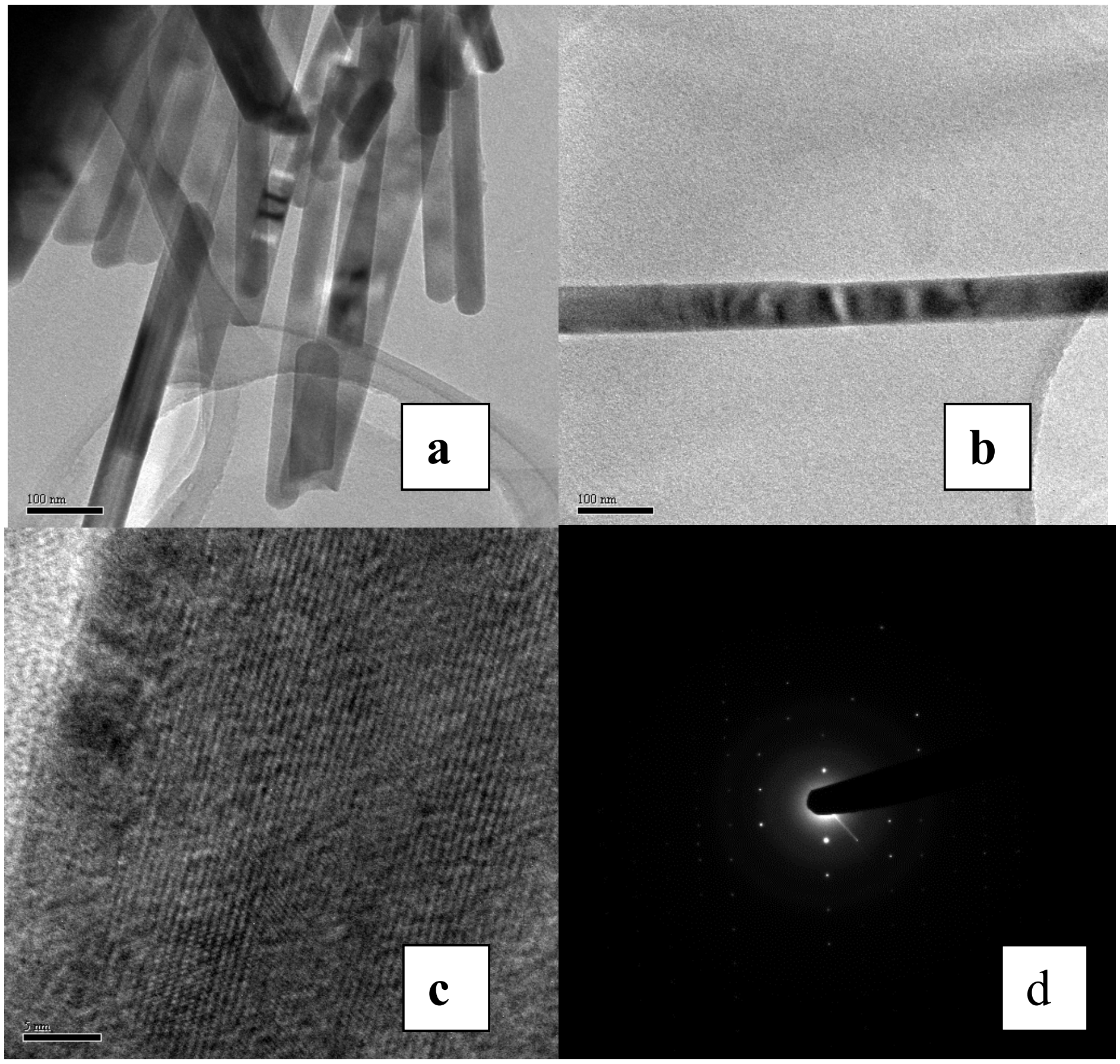

Figure 2; Rao et al

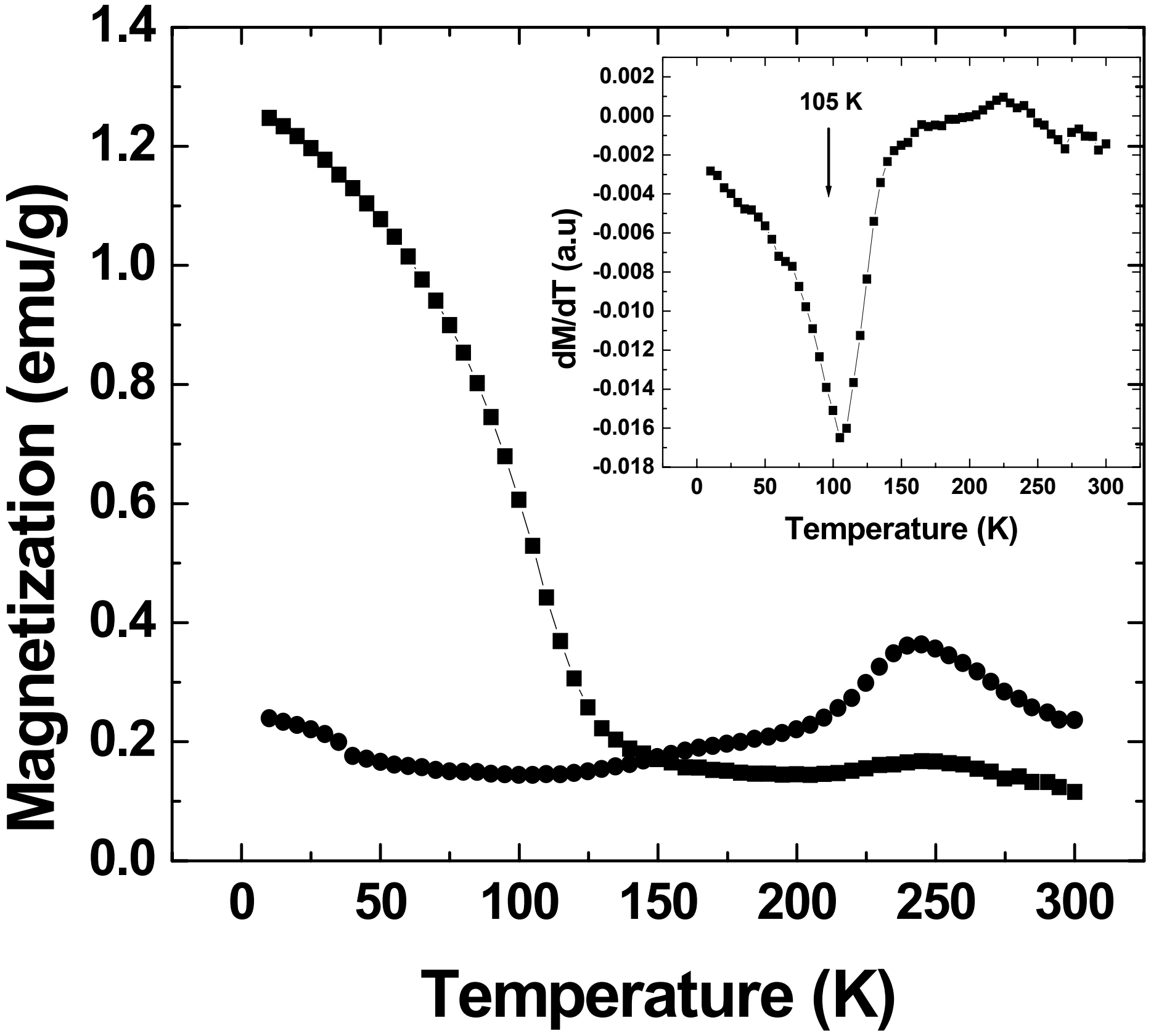

Figure 3; Rao et al

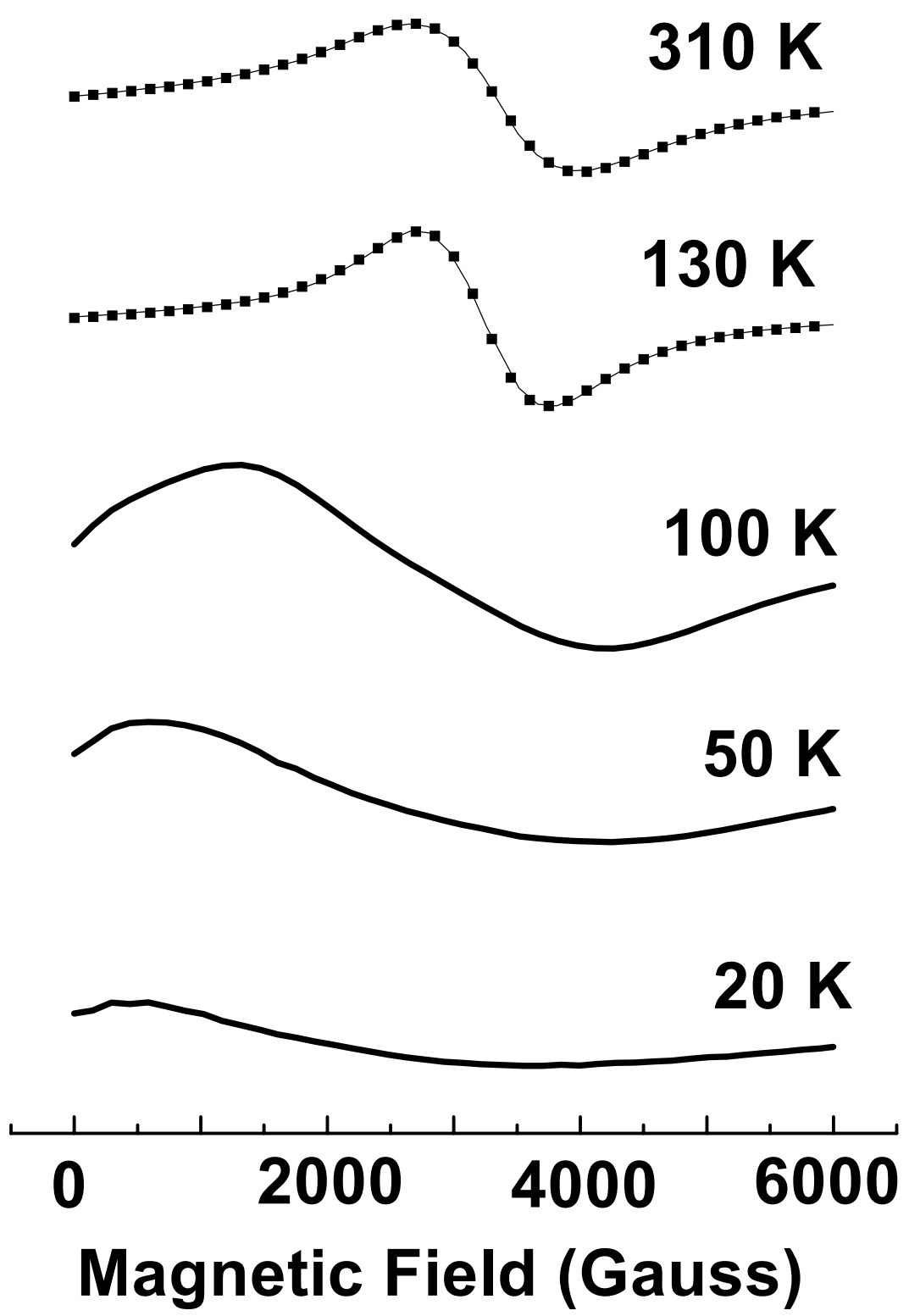

Figure 4; Rao et al